\documentclass[reprint,prd,12pt,superscriptaddress,onecolumn,nofootinbib]{revtex4-2}
\usepackage{hyperref}
\usepackage{ifpdf}
\usepackage{subfig}
\usepackage[normalem]{ulem}
\usepackage{amsmath,amssymb,amsfonts,gensymb}
\usepackage{epsf}
\usepackage{rotating}
\usepackage{slashed}
\usepackage{mathtools}
\usepackage{nicefrac}
\usepackage{color,xcolor,pstricks}
\usepackage{fancyhdr}
\usepackage{lineno}
\usepackage{braket}
\usepackage{multirow}
\usepackage{empheq}
\usepackage{graphics}
\usepackage{bm}
\def\sss{\scriptscriptstyle}

\definecolor{bv}{rgb}{0.54, 0.17, 0.89}
\allowdisplaybreaks

\def\bar {\overline}

\def \CP {$C\!P$~}

\def\beq{\begin{equation}}
\def\eeq{\end{equation}}
\def\bea{\begin{eqnarray}}
\def\eea{\end{eqnarray}}
\def\barr{\begin{array}}
\def\earr{\end{array}}
\def\nn {\nonumber}

 \usepackage{color}

 \usepackage[normalem]{ulem}

 \definecolor{darkgreen}{cmyk}{1,0,1,0.4}

\def\com2#1{\textcolor{red}{\textit{#1}}}

\begin{document}

\title{Extracting $\gamma$ from \CP violating decays of bottom baryons}


\author{Shibasis Roy} 
\email{shibasis.cmi@gmail.com}
\affiliation{Department of Physics, University of Calcutta,\\ 
92 Acharya Prafulla Chandra Road, Kolkata 700009, India}

\author{N.~G.~Deshpande}
\email{desh@uoregon.edu}
\affiliation{Institute for Fundamental Science, University of Oregon,
		Eugene, Oregon 94703, USA}

\author{Anirban Kundu}
\email{akphy@caluniv.ac.in}
\affiliation{Department of Physics, University of Calcutta,\\ 
92 Acharya Prafulla Chandra Road, Kolkata 700009, India}

\author{Rahul Sinha}
\email{sinha@imsc.res.in}
\affiliation{University of Hawaii at Manoa, Honolulu, HI 96822, USA}
\affiliation{The Institute of Mathematical Sciences, Taramani,
		Chennai 600113, India}

\begin{abstract}
The observation of \CP violation in meson decays is a testament to the 
Cabibbo-Kobayashi-Maskawa (CKM) quark mixing paradigm, and an integral
part of the Standard Model (SM). The SM also predicts the existence of \CP violation in baryon decays that is yet to be
observed.  A critical test of the SM requires that \CP violation be measured in baryon decays as well, in order to verify 
that it agrees with the measurement using meson decays. In this paper, we propose a modification to the recently
proposed method to measure \CP violating phase $\gamma$ in $b$-baryons, using interference arising implicitly due to
Bose symmetry considerations of the decay amplitudes.
\end{abstract}
 
\maketitle
\section{Introduction}

In the Standard Model, the observed \CP violation in weak interactions is attributed to the complex phase in the CKM 
quark mixing matrix. In order to observe \CP violation, the decay amplitude must
have two contributions with different strong and weak phases. In the case of neutral $B$-mesons,
mixing between particle and antiparticle allows for two
distinct amplitudes to interfere, with one amplitude
corresponding to direct decay and the other to decay via
mixing. This results in the well-known time-dependent \CP violation involving $B$-mesons. However, baryon-antibaryon oscillation is forbidden by baryon number conservation disallowing such time-dependent
mixing and consequent interference of two amplitudes. It is nevertheless still possible to observe \CP
violation in baryon decays within the SM and is being actively pursued in experiments~\cite{Aaltonen:2011qt,Aaltonen:2014vra, Aaij:2016cla, Aaij:2018lsx,
Aaij:2018tlk, Aaij:2019rkf,Aaij:2019mmy,Aaij:2019pqz,Cerri:2018ypt,LHCb:2021enr}. Since sizable weak phase within the SM appears in $b\to u$ transition, weak decays of the $b$-baryons are ideal to explore \CP violation in the baryonic 
system~\cite{Lu:2009cm,Gronau:2013mza,Hsiao:2014mua,Zhu:2016bra,Hsiao:2017tif}. A measurement of the weak phase $\gamma$ in $b$-baryons involves two interfering amplitudes arising from decay to the same final state via
intermediate $D^{0}$ and $\overline{D^{0}}$ decays~\cite{Giri:2001ju,Geng:2022osc}. This way of generating interference between two decay amplitudes is analogous to the one used
in $B$-meson decays to measure $\gamma$~\cite{Bigi:1988ym,Gronau:1990ra,Gronau:1991dp,Atwood:1996ci,Atwood:2000ck,Gronau:2002mu,Aleksan:2002mh,Grossman:2002aq,Giri:2003ty,Atwood:2003mj,Fleischer:2003aj,Fleischer:2003ai,Belle:2004bbr,Bondar:2005ki,Bondar:2008hh,Gershon:2008pe,Gershon:2009qc,Brod:2013sga,Craik:2017dpc}.


Recently, a proposal to measure weak phase $\alpha$ relying on processes specific to baryons was outlined in Ref.~\cite{Sinha:2021mmx}. In this paper, we extend that approach to measure the weak phase $\gamma$ in baryon decays using interference arising due to Bose symmetry considerations of the decay amplitudes. We start by exploring the weak
decays of the antitriplet $b$-baryons which have been
studied earlier in Refs.~\cite{He:2015fwa,He:2015fsa,Roy:2019cky,Roy:2020nyx} and where large \CP
violation may be expected~\cite{Lu:2009cm,Hsiao:2014mua,Zhu:2016bra,Hsiao:2017tif}. In addition, the
criterion of re-parametrization invariance~\cite{London:1999iv,Imbeault:2006nx} must be satisfied. This
re-parametrization condition is severely restrictive for the weak decay of
$b$-baryon and we find that even when the interference criterion due to
Bose symmetry holds, measurement of the weak phase is not possible in most cases
since re-parametrization invariance of the amplitude does not hold. After considering
all the weak decay modes of the $b$-baryons, we identify one mode namely,
$\Xi_b\to \Xi^\prime(1530)\pi$, where it is possible to measure the \CP 
violating weak phase $\gamma$. It is also expected that $\Xi_b\to \Xi^\prime(1530)\pi$ will have higher yield in comparison 
to $\Xi_b\to \Sigma^\prime(1385)\pi$ considered in Ref.~\cite{Sinha:2021mmx}, 
resulting in improved sensitivity to the weak phase. 

The overview of the method to extract $\gamma$ is presented in Sec~\ref{sec2}. The details of the method is indicated in Sec~\ref{sec3}. We finally conclude in Sec~\ref{sec4}.

\section{Overview}
\label{sec2}
\begin{figure}[h]
\includegraphics[scale=0.35]{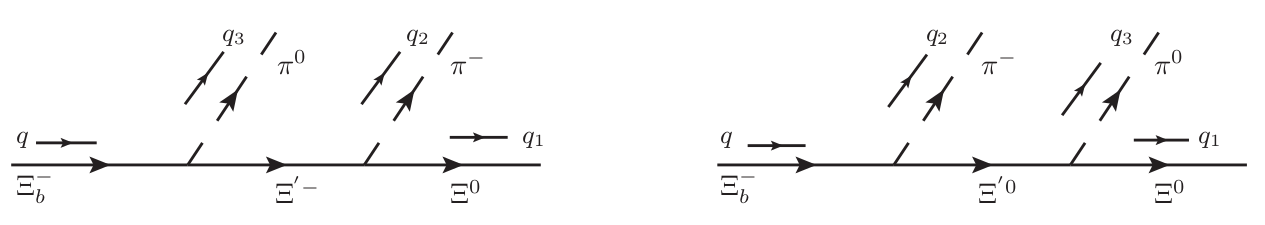}
\caption{Feynman diagrams contributing to the quasi two body decay  
$\Xi_b^-(q)\to \Xi^0(q_1)\pi^-(q_2)\pi^0(q_3)$ via the 
$J^P=\tfrac{3}{2}^+$,
$\Xi^\prime(1530)$ baryon resonance.}
\label{Fig:diagram1}
\end{figure}

For a measurement of weak phase using $b$-baryons, we focus on the decay $\Xi_b\to
\Xi^\prime\pi$, where $\Xi^\prime$ is the $J^P=\tfrac{3}{2}^+$ decuplet 
baryon resonance
corresponding to $\Xi^\prime(1530)$. We need to consider the decays
$\Xi_b^-\to\Xi^{\prime 0}\pi^-$ and $\Xi_b^-\to\Xi^{\prime -} \pi^0$
depicted in Fig.~\ref{Fig:diagram1}. 

The subsequent strong decays of the $\Xi^{\prime 0}\to \Xi^0\pi^0$
and $\Xi^{\prime -}\to \Xi^0\pi^-$ result in the identical final state
$\Xi_b^-\to \Xi^0\pi^-\pi^0$ for both the decay modes. The
$\Xi^0\pi^-\pi^0$ Dalitz plot would carry the imprints of these two decays and have
special properties under exchange of the two-pions which are identical 
bosons under isospin symmetry. While the two decay modes do not effectively interfere on the Dalitz plot, there is an 
implicit interference between these two modes from Bose symmetry correlations~\cite{Sinha:2021mmx,Sinha:2011ky}. 
This point is further illustrated below. The isospin state of the two-pions
$\pi^-\pi^0$ can either be $|I,I_3\rangle = \left|1,-1\right>$ or 
$\left|2,-1\right>$. Since the two pions are identical under isospin, their
total wavefunction must be symmetric. This necessitates that the two pions in
the odd isospin state $\left|1,-1\right>$ to be anti-symmetric under spatial
exchange, whereas, the two pions belonging to the even isospin state $\left|2,-1\right>$
be symmetric under spatial exchange. The $\Xi^0$ is an isospin
$\left|\frac12,\frac12\right>$ state, hence, the total isospin of the final
$\Xi^0\pi^-\pi^0$ state can be $\left|\frac52,-\frac12\right>$,  $\left|\frac32,-\frac12\right>$ or $\left|\frac12,-\frac12\right>$.
Isolating the $\pi^-\pi^0$ symmetric state which has isospin
$\left|2,-1\right>$, would thus be equivalent to isolating the $\Delta I=1$
contribution to the decay, which can arise from the tree or electroweak penguin diagram but not from the QCD penguin 
diagram. The $\Delta I=1$ contribution no longer carries the pure $b\to u$ weak phase due to the presence of electroweak 
penguin (EWP) which has a different weak phase. Fortunately, the EWP contribution can be parametrized entirely in terms 
of the tree amplitude using a SM relation based on SU(3)-flavor symmetry~\cite{Deshpande:1994pw,Gronau:1998fn,Neubert:1997wb} and thus it is still possible to extract the
$b\to u$ weak phase from the $\Delta I=1$ contribution.

\section{Methodology}
\label{sec3}
\begin{figure}[h]
\centering 
\includegraphics[scale=0.3]{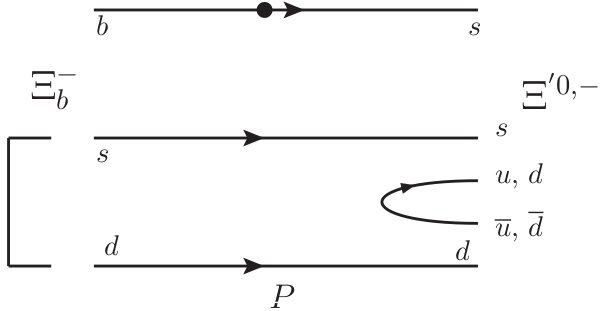} 
\includegraphics[scale=0.3]{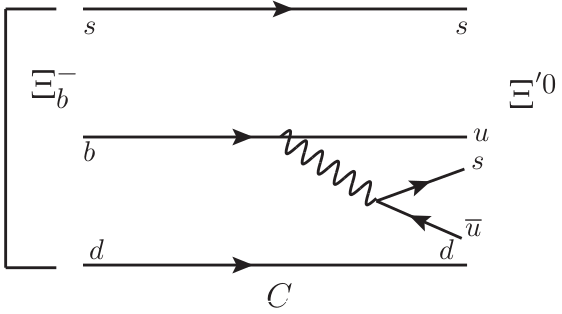} 
\caption{Topological diagrams
contributing to decay $\Xi_b^-\to \Xi^\prime \pi$ decays. The blob on the 
$b\to s$
transition in the diagram on the left corresponds to the $b\to s$ penguin. } 
\label{Fig:Topologies}
\end{figure}

The topological diagram contributions to the decays $\Xi_b^-\to\Xi^{\prime 0}\pi^-$ and $\Xi_b^-\to\Xi^{\prime -} 
\pi^0$ are shown in Fig.~\ref{Fig:Topologies}. The $s$ and $d$ quarks in the initial $\Xi_b^{-}$ baryon
are antisymmetric, as they belong to the $\bar{3}$ state of SU(3)-flavor, while the $J^P=\frac32$ 
$\Xi^\prime$ baryon final state has a completely symmetric  SU(3)-flavor wavefunction. 
Therefore the only allowed flavor-flow topologies are penguin ($P$) and color-suppressed tree ($C$), 
which ensures that the initial state $s$ and $d$ quark together do not end up in the decuplet baryon.  


The decay $\Xi_b^-\to\Xi^{\prime 0}\pi^-$ gets contributions from the color-suppressed tree and the
QCD-penguin diagrams, the corresponding amplitudes being denoted by $C$ and $P_{tc}$ respectively.
Only the QCD-penguin amplitude $P_{tc}$ contributes to the decay  $\Xi_b^-\to\Xi^{\prime -}\pi^0$. 
Here $P_{tc}=P_{t}-P_{c}$ and contribution from $P_{uc}$ is subsequently ignored. The EWP
amplitude $P_{EW}$, however, has a non-negligible impact on both these decays and therefore needs to be taken 
into account. The expressions of the decay amplitudes in terms of the topological amplitudes are given by,
\begin{eqnarray}
{\cal A}(\Xi_{b}^{-}\to  \Xi^{'0}\pi^{-}) &=& -\frac{1}{\sqrt{6}} \left[ V_{ub}V_{us}^{*}\, C+V_{tb}V_{ts}^{*}\,
\left(P_{tc}+\frac{2}{3}P_{EW}\right) \right]\,,\nonumber\\
{\cal A}(\Xi_{b}^{-}\to \Xi^{'-}\pi^{0}) &=&\frac{1}{2\sqrt{3}}V_{tb}V_{ts}^{*}\, \left(P_{tc}-\frac{1}{3}P_{EW}\right)\,.
\end{eqnarray}
A further simplification occurs when the tree and electroweak penguin part of the hadronic charmless $\vert\Delta S\vert=1$ dim-6 effective Hamiltonian~\cite{Ciuchini:1993vr,Buchalla:1995vs,Buras:1998raa} gets related in the limit of SU(3)-flavor symmetry. The $\mathbf{15}$ part of the tree and electroweak penguin effective Hamiltonian 
obeys the following relation~\cite{Gronau:1998fn}, 
\begin{align}
\mathcal{H}^{\text{EWP}}_{\mathbf{15}}(\Delta 
S=-1)&=-\frac{3}{2}\, \frac{V_{tb}V_{ts}^{*}}
{V_{ub}V_{us}^{*}}\,\frac{(C_{9}+C_{10})}{(C_{1}+C_{2})}\mathcal{H}^{\text{T}}_{\mathbf{15}}(\Delta
S=-1)
\end{align}
where $C_{1}, C_{2}$ and $C_{9},C_{10}$ are the Wilson coefficients corresponding to the tree and electroweak penguin 
operators. The topological amplitude $C$ is expressible exclusively in terms of $\mathbf{15}$ part of the tree 
Hamiltonian and is given by a single $SU(3)$-reduced amplitude~\cite{Roy:2020nyx}.    
In SM, an algebraic relation between the $P_{EW}$ and $C$ is established by noting that the same SU(3)-reduced amplitude appears in both cases. The abovementioned relation is given by
\begin{eqnarray}
\label{su3rel}
P_{EW}=\kappa C, \qquad \kappa=-\frac{3}{2}\, \frac{C_{9}+C_{10}}{C_{1}+C_{2}}\,.
\end{eqnarray}
With the help of this relation, one may rewrite the decay amplitudes as 
\begin{eqnarray}
\begin{split}
\mathcal{A}(\Xi_{b}^{-}\to \Xi^{'0}\pi^{-})&=-\frac{1}{\sqrt{6}}\, \left[\vert V_{ub}V_{us}^{*}\vert e^{-i\gamma}\, C+
V_{tb}V_{ts}^{*}\, \left(P_{tc}+\frac{2}{3}\kappa C\right) \right]\,,\\
\mathcal{A}(\Xi_{b}^{-}\to \Xi^{'-}\pi^{0})&=\frac{1}{2\sqrt{3}}V_{tb}V_{ts}^{*}\, \left(P_{tc}-\frac{1}{3}\kappa C\right)\,.
\end{split} 
\end{eqnarray} 
It is worthwhile to note that the inclusion of the electroweak penguin amplitude does not bring an additional weak phase to the $\Xi_b^-\to\Xi^{\prime -}\pi^0$ decay amplitudes. Therefore, the weak phase present in $\Xi_b^-\to\Xi^{\prime -}\pi^0$ decay amplitudes is still uniquely defined. This fact plays an important role in determination of the \CP violating weak phase $\gamma$.

The matrix element for the weak-decay $\Xi_b^-\to \Xi^{'(-,0)}\pi^{(0,-)}$
is given by
\begin{equation}
{\cal M}(\Xi_{b}^{-}\to\Xi^{'}\pi)=-i q^\pi_\mu\, 
\bar{u}^\mu_{\sss\Xi^{'}}(a+b\gamma_5)u_{\sss\Xi_b}\,,
\end{equation}
where, $u^\mu_{\sss\Xi^{'}}$ is the Rarita-Schwinger spinor for the
spin-$\frac32$ decuplet carrying a Lorentz index, $q^\pi_\mu$ is the
momentum of the $\pi$, and $u_{\sss\Xi_b}$ is the spinor of the $\Xi_{b}^{'}$. The
two coefficients $a$ and $b$ depend on the CKM elements and flavor structure
specific to the decay mode. It may be noted that $a$ and $b$ are related to the
$p$ and $d$ wave decay amplitudes respectively. The $\Xi^{'}$
baryon subsequently decays via strong-interaction to a $\Xi^{0}$-baryon and a 
pion with
the matrix element for the decay being given by,
\begin{equation}
{\cal M}(\Xi^{'}\to \Xi^0\pi)=i g_{\sss \Xi^{'} \Xi^{0}\pi}  q^\pi_\mu\, 
\bar{u}_\Xi\, u^\mu_{\Xi^{'}}\,,
\end{equation}
where, $g_{\sss \Xi^{'} \Xi^{0}\pi}$ is the invariant coupling for the decay. The
decay for the conjugate modes, $\overline{\Xi_b}^{+}\to\overline{\Xi}^{'(+,0)}\pi^{(0,+)}$ can be 
described analogously where the new decay coefficients $\overline{a}$ and $\overline{b}$
are related to $a$ and $b$ by a sign flip of the weak phases. The 
propagator of the $\Xi^{'}$ baryon has the form \cite{Choi:1989yf}
\begin{align}
\Pi^{\mu\nu}(k)&=-\frac{(\slashed{k}+m)}{(k^2-m^2+im \Gamma)}
\left(g^{\mu\nu} -\frac{2}{3} \frac{k^\mu k^\nu}{m^2}  
-\frac{1}{3} \gamma^\mu \gamma^\nu +\frac{1}{3m}
(k^\nu\gamma^\mu-k^\mu\gamma^\nu)\right)\,,
\end{align}
corresponding to that of a spin-$\frac32$ fermion with the four-momentum
$k$, mass $m$ and width $\Gamma$.
The subsequent strong decays of $\Xi^{'0}$ and $\Xi^{'-}$ to $\Xi^{0}\pi^{0}$ and $\Xi^{0}\pi^{-}$ respectively result in identical final state $\Xi^{-}_{b}\to \Xi^{0}\pi^{0}\pi^{-}$ for both the decay modes. The interference between these two modes would arise implicitly from Bose symmetry correlations between the $\pi^{0}\pi^{-}$ pair. 

The matrix element for the two-step decay $\Xi_b^-(q)\to
\Xi^{\prime -}[\to \Xi^0(q_1)\pi^-(q_2)] \pi^0(q_3)$ is given by
\begin{multline}
\label{eq:Mu}
{\cal M}_u={\cal M}\Big(\Xi_b^-(q)\to \Xi^{\prime-}\big[\to
\Xi^0(q_1)\pi^-(q_2)\big] \pi^0(q_3)\Big)=
g_{\sss \Xi^\prime \Xi \pi} 
\bar{u}(q_1)(a^{\sss-}+b^{\sss-}\gamma_5) 
\Pi^{\mu\nu}(q_{12}) 
u(q)\;q_3^\mu q_2^\nu\,,
\end{multline}
where $q_{ij}=q_i+q_j$ and $m$ is the mass of $\Xi^\prime$ resonance. 
Similarly, the two-step decay $\Xi_b^-(q)\to\Xi^{\prime 0}[\to \Xi^0(q_1)
\pi^0(q_3)] \pi^-(q_2)$ is given by
\begin{multline}
\label{eq:Mt}
{\cal M}_t={\cal M}\Big(\Xi_b^-(q)\to \Xi^{\prime 0}\big[\to
\Xi^0(q_1)\pi^0(q_3)\big] \pi^-(q_2)\Big)=
g_{\sss \Xi^\prime \Xi \pi}\bar{u}(q_1)(a^{\sss0}+b^{\sss0}\gamma_5) 
\Pi^{\mu\nu}(q_{13})
u(q)\;q_2^\mu q_3^\nu\,.
\end{multline}
The matrix element for the decay $\Xi_b^-\to \Xi^\prime[\to 
\Xi\pi] \pi$ is the sum of the matrix elements ${\cal M}_u$ and ${\cal M}_t$ such that it is Bose symmetric under the exchange of the two pions and can be expressed in a manifestly symmetric form,
\begin{multline}
{\cal M}\left(\Xi_b^-\to \Xi^\prime\left[\to 
\Xi\pi\right] \pi\right)= g_{\sss \Xi^\prime \Xi \pi} \bar{u}(q_1)\,
\left[(A_{e}+B_{e}\gamma_5)\left(\Pi^{\nu\mu}(q_{12}) +\Pi^{\mu\nu}(q_{13})\right)\right.
\\+\left. (A_{o}+B_{o}\gamma_5)  
\left(\Pi^{\nu\mu}(q_{12})-\Pi^{\mu\nu}(q_{13})\right)\right]\, u(q) q_2^\mu 
q_3^\nu\,,
 \label{eq:Bose}
\end{multline}
where $A_e$, $B_e$ and $A_o$, $B_o$ are the  even and odd parts of the 
amplitude under the 
exchange of the two pions and are given by
\begin{align}
A_{e,o}=&(a^{\sss-}\pm a^{\sss0})/2 \,, \nn \\
B_{e,o}=&(b^{\sss-}\pm b^{\sss0})/2\,.
\end{align}
From isospin symmetry that is preserved in strong decays, one obtains that the decay amplitudes of subsequent strong decays $\Xi^{'0}\to \Xi^{0}\pi^{0}$ and $\Xi^{'-}\to \Xi^{0}\pi^{-}$ follow the relation,
\begin{align}
\mathcal{A}(\Xi^{'0}\to \Xi^{0}\pi^{0})=\frac{1}{\sqrt{2}}\, \mathcal{A}(\Xi^{'-}\to \Xi^{0}\pi^{-})\,.
\end{align}
Factoring this into the definition of the decay parameters $a^{0},\, b^{0}$ and $a^{-},\, b^{-}$ for the two primary decays $\Xi_{b}^{-}\to \Xi^{'0}\pi^{-}$ and $\Xi_{b}^{-}\to \Xi^{'-}\pi^{0}$, one finds,
\begin{eqnarray}
\begin{split}
a^{-}&=\frac{\lambda_{t}}{2\sqrt{3}}\, \left(P^{p}-\frac{1}{3}\kappa C^{p}\right)\, \\
a^{0}&=-\frac{1}{2\sqrt{3}}\, \left[ \lambda_{u} e^{-i\gamma}C^{p}+\lambda_{t}\left(P^{p}+
\frac{2}{3}\kappa C^{p}\right)\right]\, \\
b^{-}&=\frac{\lambda_{t}}{2\sqrt{3}}\, \left( P^{d}-\frac{1}{3}\kappa C^{d}\right)\,, \\
b^{0}&=-\frac{1}{2\sqrt{3}}\, \left[ \lambda_{u}  e^{-i\gamma} C^{d}+\lambda_{t}\left( P^{d}+\frac{2}{3}\kappa C^{d}
\right)\right]\,,
\end{split}
\end{eqnarray}
where $P^{p}$ and $C^{p}$ are related to the topological amplitudes contributing to the $p$-wave, with analogous 
definitions for the $d$-wave, and $\lambda_{u}=\vert V_{ub}V_{us}^{*}\vert$, $\lambda_{t}=\vert V_{tb}V_{ts}^{*}\vert$ are 
shorthand for the combinations of the CKM elements. 

These even and odd parts can now be expressed by,
\begin{eqnarray}
\label{eq:xyz}
\begin{split}
A_{e}&=-\frac{\lambda_{u} C^{p}}{4\sqrt{3}}(e^{-i\gamma}+\tilde{\kappa})=-(w_{p} e^{-i\gamma}+\tilde{\kappa}w_{p})\,,\\
A_{o}&=\frac{1}{4\sqrt{3}}\left[\lambda_{u}e^{-i\gamma} C^{p}+\lambda_{t}\left
(2P^{p}+\frac{1}{3}\kappa C^{p}\right)\right]=(w_{p} e^{-i\gamma}+z_{p}e^{i\delta_{p}})\,,\\
B_{e}&=-\frac{\lambda_{u} C^{d}}{4\sqrt{3}}(e^{-i\gamma}+\tilde{\kappa})=-(w_{d} e^{-i\gamma}+\tilde{\kappa}w_{d})\,,\\
B_{o}&=\frac{1}{4\sqrt{3}}\left[\lambda_{u}e^{-i\gamma} C^{d}+\lambda_{t}\left(
2 P^{d}+\frac{1}{3}\kappa C^{d}\right)\right]=(w_{d} e^{-i\gamma}+z_{d}e^{i\delta_{d}})\,,
\end{split}
\end{eqnarray}
where 
\begin{equation}
w_{p}=\frac{\lambda_{u}C^{p}}{4\sqrt{3}}\,,\ \ \ 
w_{d}=\frac{\lambda_{u}C^{d}}{4\sqrt{3}}\,,\ \ \ 
z_{p}=\frac{\vert \lambda_{t}(2P^{p}+\frac{1}{3}\kappa C^{p})\vert}{4\sqrt{3}}\,,\ \ \ 
z_{d}=\frac{\vert \lambda_{t}(2P^{d}+\frac{1}{3}\kappa C^{d})\vert}{4\sqrt{3}}\,.
\end{equation}
$\delta_{p}$, $\delta_{d}$ are the strong phase differences between the penguin and tree amplitudes for the $p$-wave and $d$-wave respectively. A numerical estimate of the parameter $\tilde{\kappa}$ is provided in terms of $\kappa$ defined in Eq.~\eqref{su3rel} and the CKM factors $\lambda_{u}$ and $\lambda_{t}$~\cite{HFLAV:2022pwe},
\begin{align}
\tilde{\kappa}=\frac{\lambda_{t}}{\lambda_{u}}\kappa=0.65\pm 0.04\,.
\end{align} 

\begin{figure}[h]
\includegraphics[scale=0.55]{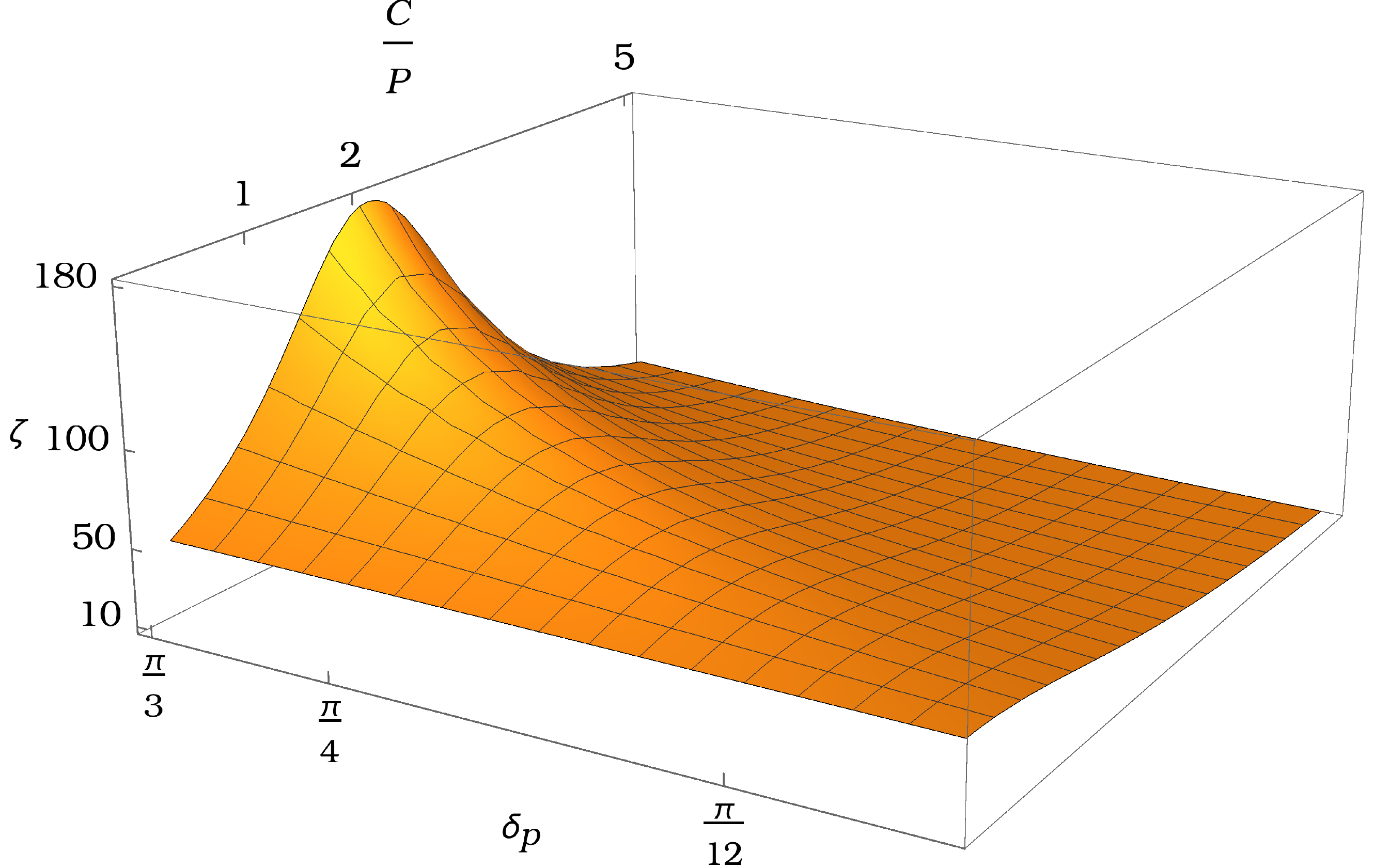}
\caption{Enhancement factor ($\zeta$) for decay rate of $\Xi_{b}^{-}\to [\Xi\pi]\pi$ in comparison to $\Xi_{b}^{-}\to [\Lambda\pi]\pi$ for $ 0.1 \leq C/P \leq 5$ and $0\leq\delta_{p}\leq\frac{\pi}{3}$
}
\label{Fig:enh}
\end{figure}

Assuming the $C/P $ ratio and the strong phase difference for the $p$-wave amplitude of $\Xi_{b}^{-}\to [\Lambda\pi]\pi$ and $\Xi_{b}^{-}\to [\Xi\pi]\pi$ remaining the same, which is a valid assumption under SU(3) flavor symmetry, 
a comparison between the decay rates is made in Fig.~\ref{Fig:enh}. It is apparent that
 $\Xi_{b}^{-}\to [\Xi\pi]\pi$ decay rate is roughly at least a factor of 10 enhanced over $\Xi_{b}^{-}\to [\Lambda\pi]\pi$ decay rate and therefore more likely to be observed first.

\begin{figure}[b]
\includegraphics[scale=0.3]{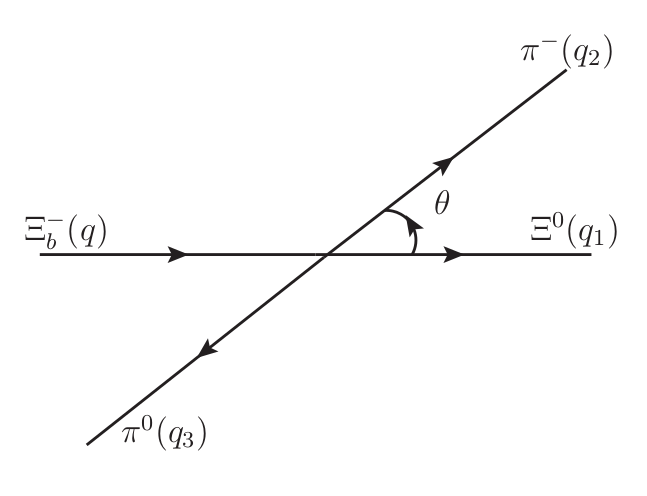}
\caption{Jackson frame in which $\Xi_b^-(q)$ decays to three bodies 
$\Xi^0(q_1)$, $\pi^-(q_2)$ and $\pi^0(q_3)$~.}
\label{Fig:Jackson}
\end{figure}

We consider the decay chain in the Gottfried-Jackson frame, Fig.~\ref{Fig:Jackson}, with $\Xi_{b}^{-}$ moving along
the $+\hat{z}$ axis, and the two pions going back to back with the 
$\pi^-(q_2)$ at an angle $\theta$ to the $\Xi^{0}(q_1)$. In this 
frame, 
$\bm{q_2}+\bm{q_3}=0$. 
We define the Mandelstam variables as usual: $s\equiv(q_2+q_3)^2=(q-q_1)^2$, 
$t\equiv(q_1+q_3)^2=(q-q_2)^2$ and $u\equiv(q_1+q_2)^2=(q-q_3)^2$, so that $t$ and $u$ 
can be written as 
\begin{align}
t&=x+y\cos\theta\,,\label{eq:t}\\
u&=x-y\cos\theta\,,
\label{eq:u}
\end{align}
where 
\begin{gather}
x=\frac{M^2+m_{\Xi}^2+2 m_\pi^2-s}{2}\,,\label{eq:x}\\
y=\frac{\sqrt{s-4 
m_\pi^2}}{2\sqrt{s}}\lambda^{\nicefrac{1}{2}}(M^2,m_{\Xi}^2,s)\label{eq:y}\,,
\end{gather}
and $\lambda(M^2,m_\Xi^2,s)=(M^4+m_\Xi^4+s^2-2 M^2 m_\Xi^2-2 M^2 
s-2 m_\Xi^2 s)$. Here, $M$ is the mass of the $\Xi_{b}^{-}$ baryon, $m_{\Xi}$ is the 
mass of the $\Xi^{0}$ and $m_\pi$ is the mass of the pions. The decay rate for the 3-body final state $\Xi^0\pi^-\pi^0$ is cast in terms of $s$
and $\cos\theta$ using Eqs.~\eqref{eq:t},~\eqref{eq:u},~\eqref{eq:x} and
\eqref{eq:y}. Under
the exchange of the two pions, $\theta \leftrightarrow \pi-\theta$. The odd
(even) part of the amplitude under the exchange of two pions must therefore be
proportional to odd (even) powers of $\cos\theta$. Bose correlations are evident in plots involving $\hat{s}=s/M^2$ and 
$\theta$ and such plots are referred to as correlation-plots henceforth.

The numerator of  the decay rate, $N_\Gamma$, for the decay mode under consideration and its \CP conjugate 
mode are worked out to have the complicated form:
\begin{equation}
N_\Gamma=\sum_{n=0}^{4}c_{n}(\hat{s}) 
\cos2n\theta+\sum_{n=0}^{3}d_{n}(\hat{s})
\cos(2n+1)\theta\,,
\label{eq:Gamma}
\end{equation}
where, all masses and momenta are normalized to $M$, the
$\Xi_b$ mass, for simplicity, and
\begin{align}
\label{eq:cn}
&c_{n}(s)=f_{n}^{\sss(1)}(\hat{s})|A_{e}|^2+ 
f_{n}^{\sss(2)}(\hat{s})|B_{e}|^2+f_{n}^{\sss(3)}(\hat{s})|A_{o}|^2 + f_{n}^{\sss(4)}(\hat{s}) 
	|B_{o}|^2 \,,
	\\ 
&d_{n}(\hat{s})=g_{n}^{\sss(1)}(\hat{s})\text{Re}(A_eA_o^*)+
	g_{n}^{\sss(2)}(\hat{s})\text{Re}(B_eB_o^*)+g_{n}^{\sss(3)}(\hat{s})\text{Im}(A_eA_o^*)
	+g_{n}^{\sss(4)}(\hat{s})\text{Im}(B_eB_o^*)\,.
\label{eq:dn}
\end{align}
The coefficients $f_{n}^{(i)}$ and $g_{n}^{(i)}$ are functions of $\hat{s}$ and
are expressed in terms of kinematic factors. For a given choice of
$\hat{s}$, $f_{n}^{(i)}$ and $g_{n}^{(i)}$ are just numbers. We henceforth drop
explicit dependence on $\hat{s}$, since our solutions are valid for all
$\hat{s}$. The numerator of the decay rate, given in Eq.~\eqref{eq:Gamma}, can be
fitted as a function of $\theta$ to obtain the coefficients $c_n$ and $d_n$. Once we have 
$c_0$, $c_1$, $c_2$ and $c_3$, it is trivial to get $|A_{e}|^2$,
$|A_{o}|^2$, $|B_{e}|^2$ and $|B_{o}|^2$. Similarly,
$\text{Re}(A_eA_o^*)$, $\text{Re}(B_eB_o^*)$, $\text{Im}(A_eA_o^*)$ and
$\text{Im}(B_eB_o^*)$ can be solved using $d_0$, $d_1$, $d_2$ and $d_3$. In
order to solve for the amplitudes and their interference, one must have a
minimum of 8-bins in $\theta$ and $\hat{s}$ involving both the resonances
contributing to the process. Experimental procedure used in an actual analysis
can be more refined. An identical analysis of the decay rate for the conjugate
process would enable us to solve for $|\overline{A}_e|^2$, $|\overline{A}_o|^2$,
$|\overline{B}_e|^2$, $|\overline{B}_o|^2$ and
$\text{Re}(\overline{A}_e\overline{A}_o^*)$,
$\text{Re}(\overline{B}_e\overline{B}_o^*)$,
$\text{Im}(\overline{A}_e\overline{A}_o^*)$ and
$\text{Im}(\overline{B}_e\overline{B}_o^*)$. Having expressed the values of these
amplitudes and their interference, our aim is to solve for the weak phase
$\gamma$, the amplitudes $w_{{p,d}}$ and $z_{{p,d}}$
defined in Eq.~\eqref{eq:xyz}, and the strong phases $\delta_{{p,d}}$.
In order to obtain the solutions for the $p$-wave parameters we define
new intermediate observables $r_i$:

\begin{align}
r_0=& |A_e|^2=|\overline{A}_e|^2=w_p^2(1+\tilde{\kappa}^{2}+2\tilde{\kappa}\cos{\gamma})\,,\nn\\
r_1=& |A_o|^2+|\overline{A}_o|^2
	=2 w_p^2+2 z_p^2+4 w_p z_p 
	\cos\delta_p \cos\gamma\,,\nn\\
r_2=& |A_o|^2-|\overline{A}_o|^2
	=-4\, w_p z_p \sin\delta_p \sin\gamma\,, \nn\\
r_3=&\text{Re}(A_eA_o^*-\overline{A}_e\overline{A}_o^*
	)=2 w_p z_p \sin\delta_p\sin\gamma\,,\nn\\
r_4=&\text{Im}(A_eA_o^*-\overline{A}_e\overline{A}_o^*)=2 w_p 
z_p \cos\delta_p\sin\gamma-2\tilde{\kappa}w_{p}^{2}\sin\gamma\,, \nn\\
r_5=&\text{Re}(A_eA_o^*+\overline{A}_e\overline{A}_o^*) 
	=-2 w_p^2  - 2 w_p z_p 
	\cos\delta_p\cos\gamma-2\tilde{\kappa}w_{p}^{2}\cos\gamma-2\tilde{\kappa}w_{p}z_{p}\cos\delta_{p}\,, \nn\\
r_6=&\text{Im}(A_eA_o^*+\overline{A}_e\overline{A}_o^*) 
	=2 w_p z_p \sin\delta_p\cos\gamma+2\tilde{\kappa}w_{p}z_{p}\sin\delta_{p}\,.
\end{align}
The intermediate observables are not all independent, and the following relations
\begin{align}
\label{int obv rel}
r_3^2 + r_4^2 + r_5^2 + r_6^2& = 2 r_0  r_1\,,\nn\\
r_3 r_5 + r_4 r_6 &= r_0 r_2\,, \nn\\
r_2=-2 r_3\,,
\end{align}
restrict the number of parameters that can be extracted.
Focusing on a solution for the weak phase $\gamma$ we find the expression,
\begin{align}
\cot \gamma= \frac{-R\pm\tilde{\kappa}\sqrt{R^{2}-\tilde{\kappa}^{2}+1}}{\tilde{\kappa}^{2}-1}
\end{align}
where $R=r_6 / r_3$.
It should be emphasized that the relations between the intermediate observables mentioned in 
Eq.~\eqref{int obv rel} must hold experimentally for an unambiguous determination of the weak phase. 
The solution for weak phase $\gamma$ using $d$-wave parameters can be
obtained similarly. It is evident that both the $p$- and $d$-
wave contributions must result in the measurement of the same weak phase
$\gamma$. We therefore obtain two independent measurements of $\gamma$ corresponding to the two partial waves.


The observed correlation plot must have a more complicated structure with several
resonances, and one may wonder if our approach to measure the weak phase would
be possible. The contribution from heavier $\Xi^\prime$ states have a 
similar decay
dynamics with the same weak phase $\gamma$, but the relevant amplitudes and
strong phases would differ. These resonances are not a cause for concern as
suitable binning cuts would easily remove their contributions without losing
relevant signal sample. Only the interference with the decay mode $\Xi_b^-\to
\Xi^0\rho^-\to \Xi^0\pi^-\pi^0$, and modes involving heavier $\rho$-like
resonances, instead of $\rho$, need a closer look. It is interesting to note
that $\Xi_b^-\to \Xi^0\rho^-$  has a different $\hat{s}$ dependence in the
overlap region and contributes only to the odd part of the amplitude. Once data
is available in several more $\hat{s}$ bins, such interference effects can
easily be isolated using the Dalitz distribution. One does not
need to impose cuts to remove the $\Xi_b^-\to \Xi^0\rho^-$ contributions.

\section{Conclusion}
\label{sec4}
In this paper, we discussed the decay $\Xi_b^- \to \Xi^0\pi^-\pi^0$, which can proceed through two intermediate
decay channels, namely, $\Xi_b^-\to \Xi^{\prime0}\pi^-$ and $\Xi_b^-\to \Xi^{\prime-}\pi^0$, and we have shown 
how Bose correlations arise from these two intermediate decays. Similar Bose correlations also arise in
the decay to conjugate final state $\overline{\Xi}_b^+\to \overline{\Xi}^0\pi^+\pi^0$. 
We provide an estimate of the enhancement in decay rate for $\Xi_{b}^{-}\to [\Xi\pi]\pi$ in comparison to 
$\Xi_{b}^{-}\to [\Lambda\pi]\pi$ which suggests a significantly larger rate and hence an improved sensitivity 
for measuring $\gamma$ in comparison to the mode previously proposed to measure $\alpha$. 
We further show that the weak phase $\gamma$ can be measured using
both even and odd contributions to the amplitudes under pion exchange and 
comparing the correlation plots for the original as well as the \CP conjugate modes. Our approach to measure \CP violation 
crucially depends on such Bose correlations. This will provide a new, and independent, way of measuring the CKM
angle $\gamma$.

\section{Acknowledgments}
S.R.\ and A.K.\ acknowledge the support through the Grant No.\ CRG/2019/000362 from the 
Science and Engineering Research Board, Government of India.


\end{document}